\documentclass[aps,showpacs,twocolumn]{revtex4-2}
\usepackage[utf8]{inputenc}
\usepackage{amsfonts}
\usepackage{amssymb}
\usepackage{amsmath}
\usepackage{graphicx}
\usepackage{epsfig}
\usepackage{subfigure}
\usepackage{appendix}
\usepackage{hyperref}
\usepackage{color}
\usepackage{mathtools}
\usepackage{bm}

\newcommand{\ii}{\mathrm{i}}
\newcommand{\ee}{\mathrm{e}}
\hypersetup{hypertex=true, colorlinks=true, linkcolor=blue, urlcolor=blue, citecolor=blue}

\begin{document}

\title{Vector chiral order and dynamical quantum phase transitions in an Ising chain with dimerized anisotropic Gamma interaction}
\author{Yu-Hong Yan}
\author{Shi-Qiao Wu}
\author{Kun-Liang Zhang}
\email[Contact author: ]{zhangkl@fosu.edu.cn}
\date{\today}

\affiliation{School of Physics and Optoelectronic Engineering, Foshan
University, Foshan, 528225, China} 

\begin{abstract}
We investigate a transverse-field Ising chain with dimerized anisotropic Gamma interactions and demonstrate that, both in the absence and presence of Ising interactions, the competition between Gamma anisotropy and the transverse field gives rise to vector chiral order and dynamical quantum phase transitions. The phase diagram comprises positive and negative vector chiral phases and a paramagnetic phase, characterized by the vector spin chirality and the vector chiral order parameter. In the absence of Ising interactions, we identify a gapless critical line that exhibits no finite-size drift, along which the spin-spin correlations and subsystem entanglement entropy display behavior distinct from that of Ising criticality. Furthermore, we find that the critical times of dynamical quantum phase transitions form periodic or aperiodic sequences, depending on the quench path. Our findings highlight how dimerized off-diagonal exchange enriches both equilibrium ordering and dynamical critical behavior in a quantum Ising chain.
\end{abstract}

\maketitle

\section{Introduction}

\label{introduction}

The transverse field Ising chain is a paradigmatic model for studying quantum phase transitions driven by the competition between Ising exchange and a transverse magnetic field \cite{Pfeuty1970}. The resulting quantum critical point separates a quantum paramagnetic phase from a magnetically ordered phase, which is ferromagnetic or antiferromagnetic depending on the sign of the exchange coupling.
Beyond this conventional setting, in recent years, bond-dependent and off-diagonal exchange interactions have attracted considerable attention as routes to unconventional quantum magnetism. Representative examples include compass-type couplings \cite{Brzezicki2007, jackeli2009mott, Sun2009, Cincio2010, You2014a, Nussinov2015, You2018, Laurell2023, Richards2024}, symmetric off-diagonal Gamma exchange \cite{SolanoCarrillo2011, Rau2014, yang2020phase, Liu2020, Soerensen2021, Zhao2022, Kheiri2024, Saito2024a, Abbasi2025, Huang2026, yan2026dynamical}, and antisymmetric Dzyaloshinskii-Moriya (DM) interactions \cite{Moriya1960, Brzezicki2007, Jafari2008, Ma2011, Rau2014, Wang2015, Cheraghi2018, Avalishvili2019, Matsumoto2021, Cao2022a, Kunimi2024}. In spin-orbit-coupled magnets, the symmetry and strength of off-diagonal exchange are closely tied to the local exchange geometry \cite{Rau2014,Moriya1960}.
Such exchange can promote noncollinear states, including incommensurate spiral order \cite{SolanoCarrillo2011, Rau2014, Li2025b, Huang2026}, while competing interactions in low-dimensional magnets can also support vector chiral order \cite{Avalishvili2019, Soerensen2021, Soerensen2023, Luo2024} and spin-nematic correlations \cite{Hikihara2008, Yang2021, Soerensen2021, luo2023spontaneous, Luo2024, Abbasi2025, yan2026dynamical}.
These developments motivate investigating how bond alternation and off-diagonal exchange reshape the phases and critical properties of the transverse field Ising chain.

Away from equilibrium, quantum spin chains provide a natural setting for exploring nonequilibrium critical phenomena under periodic driving or following sudden quenches \cite{Heyl2013, Heyl2018, Yang2019, Mazza2019, Lerose2020, zhang2020dynamic, Ippoliti2021, Morris2021, Zhang2021a, Ghosh2022, Pomponio2022, Hansen2022, Cao2022a, zhang2024magnetic, zhang2024two, zhang2024magnetization, Cao2024, shi2024emerging, Cheng2025a, xiong2025resonant, yan2026dynamical}.
Dynamical quantum phase transitions (DQPTs) are characterized by nonanalyticities in the Loschmidt return-rate function at critical evolution times in the thermodynamic limit \cite{Heyl2013, Heyl2018, Yang2019, Cao2022a, Cao2024, shi2024emerging, cao2026tailoring}.
Although DQPTs are closely connected to quenches across the equilibrium critical point in the conventional transverse-field Ising chain, this correspondence is not universal \cite{Vajna2014}. It is therefore natural to ask how additional off-diagonal exchange channels and chiral correlations influence dynamical criticality, particularly whether different quench paths lead to periodic or aperiodic sequences of critical times.

Motivated by the aforementioned spin-orbit-coupled quantum magnets and low-symmetry exchange pathways, we introduce a dimerized anisotropic Gamma interaction into the transverse field Ising chain. This interaction couples the $x$ and $y$ spin components on neighboring sites, with the parameter $\alpha$ controlling the relative strengths of the two off-diagonal exchange channels.
Their alternation between odd and even bonds mimics a dimerized local environment.
For the bond pattern considered here, the interaction can equivalently be decomposed into a uniform symmetric Gamma exchange and a staggered DM interaction, providing a route to study the interplay among dimerization, spin-orbit-induced anisotropy, vector chirality, and Ising criticality. 
We investigate both the equilibrium phase diagram and the dynamics following sudden quenches, in the absence and presence of Ising interactions.
Using spin-spin correlation functions and the vector chiral order parameter, we identify positive and negative vector chiral phases, together with a paramagnetic phase. In the absence of Ising interactions, we uncover a gapless critical line whose location exhibits no finite-size drift, along which the spin correlations and subsystem entanglement entropy display behavior distinct from that associated with Ising criticality.
Furthermore, we demonstrate that the critical times of DQPTs form periodic or aperiodic sequences depending on the quench path.
These findings highlight how dimerized off-diagonal exchange enriches both equilibrium ordering and dynamical critical behavior in a quantum Ising chain.

The rest of this paper is organized as follows: In Sec. \ref{sec_model}, we
introduce  and diagonalize the Hamiltonian of the Ising chain with dimerized
anisotropic Gamma interaction under a transverse magnetic field, and obtain
the phase diagram defined by the energy gap. In Sec. \ref{sec_correlations},
we present the results of spin-spin correlation functions and vector chiral order parameter; Discuss the behavior of the correlation function and subsystem entanglement entropy on different critical lines. In Sec. \ref{sec_DQPTs}, we investigate the DQPTs through the rate function of the system following different types of quench path. Finally, we discuss and conclude our findings in Sec. \ref{sec_conclusion}.

\section{Model and exact solution}

\label{sec_model} 

The model considered here is an Ising chain with dimerized anisotropic Gamma interaction under a transverse magnetic field, the Hamiltonian of which can be written as 
\begin{equation}
\begin{aligned} \hat{H}=&J\sum_{i=1}^{N} \hat{\sigma}_i^x\hat{\sigma}_{i+1}^x+
h\sum_{i=1}^{N}\hat{\sigma}_i^z\\ &+\sum_{i=1}^{N} \left(
J_i^{xy}\hat{\sigma}_i^x\hat{\sigma}_{i+1}^y + J_i^{yx}\hat{\sigma}_i^y\hat{\sigma}_{i+1}^x \right)
, \end{aligned}
\label{eq_Hs}
\end{equation}
where the strength of anisotropic Gamma couplings are taken as $J_i^{xy}=%
\mathit{\Gamma}$, $J_i^{yx}=\alpha\mathit{\Gamma}$ for odd $i$, and $%
J_i^{xy}=\alpha\mathit{\Gamma}$, $J_i^{yx}=\mathit{\Gamma}$ for even $i$.
The first term is the antiferromagnetic Ising coupling with strength $J$, and the second term is the transverse field with strength $h$. We assume that
the length of the chain $N$ is even, and periodic boundary condition is imposed.

To diagonalize the Hamiltonian, we first introduce the Jordan-Wigner transformation \cite{jordan1928paulische}
\begin{equation}
\begin{aligned} 
\hat{\sigma}_j^x &= \prod_{l<j} \left(1-2\hat{c}_l^\dagger \hat{c}_l\right)
\left(\hat{c}_j^\dagger+\hat{c}_j\right), \\ \hat{\sigma}_j^y &= \mathrm{i} \prod_{l<j}
\left(1-2\hat{c}_l^\dagger \hat{c}_l\right) \left(\hat{c}_j^\dagger-\hat{c}_j\right), \\
\hat{\sigma}_j^z &= 2\hat{c}_j^\dagger \hat{c}_j-1, 
\end{aligned}
\end{equation}
and obtain the fermionic representation of the Hamiltonian 
\begin{equation}
\begin{aligned} 
\hat{H}= \sum_{i=1}^{N}& \Bigl\{ \left[J+\mathrm{i}(-1)^i(1-\alpha)\mathit{\Gamma}\right] \hat{c}_i^\dagger
\hat{c}_{i+1} \\ 
&+\left[J+\mathrm{i}(1+\alpha)\right] \hat{c}_i^\dagger \hat{c}_{i+1}^\dagger +\mathrm{H.c.} \Bigr\} \\ 
&+2h\sum_{i=1}^{N}\hat{c}_i^\dagger \hat{c}_i -Nh . 
\end{aligned}
\label{eq_Hf}
\end{equation}
For simplicity, here we neglect the fermion-parity-dependent factor generated by the Jordan-Wigner
transformation under periodic boundary conditions. This factor only
determines whether the allowed momenta are integer- or
half-integer-quantized, and does not affect the dispersion relation in the
thermodynamic limit. Hamiltonian in Eq. (\ref{eq_Hf}) is a quadratic fermionic chain with staggered 
hopping and nearest-neighbor $p$-wave pairing.

Then we taking the discrete Fourier transformation 
\begin{equation}
\hat{c}_{2l-1}=\frac{1}{\sqrt{N/2}}\sum_{k}e^{\ii kl}\hat{a}_{k},\quad  \hat{c}_{2l}=\frac{1}{%
\sqrt{N/2}}\sum_{k}e^{\ii kl}\hat{b}_{k},
\end{equation}%
where $k\in \lbrack -\pi ,\pi )$ is the momentum associated with the
two-site unit cell. The common phase factor can be eliminated by the gauge transformation $\hat{a}_{k}=e^{\ii k/2}
\widetilde{\hat{a}}_{k}$ and  $\hat{b}_{k}=e^{\ii k/2}
\widetilde{\hat{b}}_{k}$. In the following, we suppress the tilde and continue to
denote the transformed operators by $\hat{a}_{k}$ and $\hat{b}_{k}$. Introducing the Nambu spinor $%
\hat{\Psi}_{k}^{\dagger }=(\hat{a}_{k}^{\dagger },\hat{b}_{k}^{\dagger },\hat{a}_{-k},\hat{b}_{-k})$, the
Hamiltonian can be written as Bogoliubov-de Gennes (BdG) form 
\begin{equation}
\hat{H}=\frac{1}{2}\sum_{k}\hat{\Psi} _{k}^{\dagger }\mathcal{H}_{\mathrm{BdG}}(k)\hat{\Psi}
_{k},
\end{equation}%
where 
\begin{equation}
\mathcal{H}_{\mathrm{BdG}}(k)=%
\begin{pmatrix}
2h & Q(k) & 0 & P(k) \\ 
Q^{\ast }(k) & 2h & -P(-k) & 0 \\ 
0 & -P^{\ast }(-k) & -2h & -Q^{\ast }(-k) \\ 
P^{\ast }(k) & 0 & -Q(-k) & -2h%
\end{pmatrix}.
\end{equation}
and 
\begin{equation}
\begin{aligned} 
Q(k) &= \left[J-\ii (1-\alpha)\mathit{\Gamma}\right](1+e^{-\ii k}), \\ 
P(k)&= \left[J+\ii (1+\alpha)\mathit{\Gamma}\right](1-e^{-\ii k}). 
\end{aligned}
\end{equation}%

The Hamiltonian can be diagonalized by introducing the Bogoliubov quasiparticle operators 
\begin{equation}
    \hat{\gamma}_{k,\lambda}
    =
    u_{k,\lambda}\hat{c}_{k,\lambda}
    -
    \ii\lambda v_{k,\lambda}
    \hat{c}_{-k,\lambda}^{\dagger}.
    \label{eq_gamma}
\end{equation}
where the band fermion operators are
\begin{equation}
    \hat{c}_{k,\lambda}
    =
    \frac{1}{\sqrt{2}}
    \left(
        \ee^{-\ii\chi}\hat{a}_k
        -
        \ii\lambda\ee^{-\ii\beta_k}\hat{b}_k
    \right),
\end{equation}
with the phase factors
\begin{equation}
\chi =\frac{\theta -\phi }{2},\qquad \beta _{k}=\chi +\frac{k}{2}+\phi +%
\frac{\pi }{2},
\end{equation}
and $\phi$, $\theta$ are determined by
\begin{equation}
\phi=\arg [J+\ii (1-\alpha )\mathit{\Gamma}],\quad \theta=\arg [J+\ii (1+\alpha )\mathit{\Gamma}].
\label{eq_phi_theta}
\end{equation}
The Bogoliubov coefficients in Eq. (\ref{eq_gamma}) are chosen as 
\begin{equation}
\begin{aligned}
& u_{k,\lambda }=\sqrt{\frac{E_{\lambda }(k)+\xi _{\lambda }(k)}{2E_{\lambda
}(k)}},\\
& v_{k,\lambda }=\mathrm{sgn}(\sin \frac{k}{2})\sqrt{\frac{%
E_{\lambda }(k)-\xi _{\lambda }(k)}{2E_{\lambda }(k)}},
\end{aligned}
\end{equation}
with $\lambda =\pm 1$, and the dispersion
\begin{equation}
\begin{aligned} E_\lambda(k) = 2&\Bigg\{ \left[ h -\lambda \sqrt{
J^2 +(1-\alpha)^2\mathit{\Gamma}^2 } \cos\frac{k}{2} \right]^2 \\
&+ \left[ J^2 +(1+\alpha)^2\mathit{\Gamma}^2 \right]
\sin^2\frac{k}{2} \Bigg\}^{1/2},\\
\xi _{\lambda }(k)=2&h-2\lambda \sqrt{J^{2}+(1-\alpha )^{2}\mathit{\Gamma} ^{2}}\cos 
\frac{k}{2}.
\end{aligned}
\end{equation}
Through the above Bogoliubov transformation, the diagonalized Hamiltonian is obtained as 
\begin{equation}
H=\sum_{k}\sum_{\lambda =\pm 1}E_{\lambda }(k)\left( \hat{\gamma} _{k,\lambda
}^{\dagger }\hat{\gamma} _{k,\lambda }-\frac{1}{2}\right) ,
\end{equation}
and the ground-state energy is therefore 
\begin{equation}
E_{\mathrm{GS}} = -\frac{1}{2} \sum_k \sum_{\lambda=\pm} E_{\lambda}(k).
\end{equation}
It can be checked that $ \hat{\gamma}_{k,\lambda}$ and $\hat{\gamma}_{k,\lambda}^{\dagger}$ obey the following anticommutation relations
\begin{equation}
	\begin{aligned}
		&\bigl\{ \hat{\gamma}_{k,\lambda},\hat{\gamma}_{k',\lambda'}^{\dagger} \bigr\}=\delta_{k,k'}\delta_{\lambda , \lambda'},\\
		&\bigl\{ \hat{\gamma}_{k,\lambda},\hat{\gamma}_{k',\lambda'} \bigr\}=\bigl\{ \hat{\gamma}_{k,\lambda}^{\dagger},\hat{\gamma}_{k',\lambda'}^{\dagger} \bigr\}=0.
	\end{aligned}
\end{equation}
The many-body ground state $\lvert\mathrm{GS}\rangle$ is defined as the vacuum of all positive-energy
Bogoliubov quasiparticles, that is, $ \hat{\gamma}_{k,\lambda}\lvert\mathrm{GS}\rangle=0$ for each $k$ and $\lambda$.
Then the ground state of the system can be written as
\begin{equation}
    \lvert\mathrm{GS}\rangle
    =
    \prod_{k>0}
    \prod_{\lambda=\pm}
    \left(
        u_{k,\lambda}
        +
        \ii\lambda v_{k,\lambda}
        \hat{c}_{k,\lambda}^{\dagger}
        \hat{c}_{-k,\lambda}^{\dagger}
    \right)
    \lvert0\rangle,
\end{equation}
where $\lvert0\rangle$ is the vacuum for the fermions, i.e., $\hat{c}_{i}\lvert0\rangle=0$.

\begin{figure}[tb]
\centering
  \includegraphics[width=0.5\textwidth]{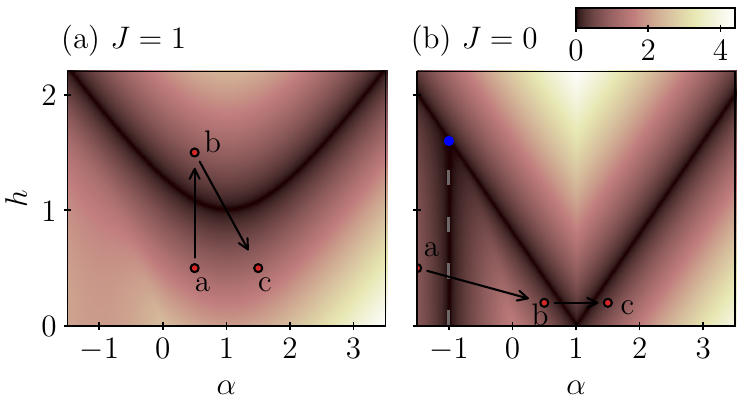}
  \caption{The contour map of the energy gap $\mathit{\Delta}$ as a function of $\alpha$ and $h$ for the cases of (a) nonzero Ising interaction $J=1$ and (b) zero Ising interaction $J=0$. The black lines are critical lines that indicate the closing of the energy gap $\mathit{\Delta}=0$, and there is an additional critical line (marked with dashed line) when $J=0$.  In both cases we take $\mathit{\Gamma}=0.8$.}
  \label{fig_gap}
\end{figure}

In Figs. \ref{fig_gap} (a) and (b), we present the contour map of the energy gap $\mathit{\Delta}$ between the ground state and first excited state as a function of $\alpha$ and $h$ in the presence and absence of Ising interactions, respectively. 
The phase boundary can be obtained by the closing condition of the energy gap, which gives $h_\mathrm{c}^2= J^2+(1-\alpha)^2\Gamma^2$. On this critical line, the gap closing momentum is $k_0=0$.  Near the gap closing momentum, we have 
\begin{equation}
    E_{+}(k)
    \simeq
    \sqrt{J^{2}+(1+\alpha)^{2}\mathit{\Gamma}^{2}} \lvert k\rvert.
\end{equation}
Therefore, the critical quasiparticle dispersion is linear and the dynamical critical exponent is $z=1$.
However, there is a additional critical line with $\alpha_\mathrm{c}=-1$ and $|h|< 2 \mathit{\Gamma}$ when taking $J= 0$ [see the dashed line in Fig. \ref{fig_gap} (b)]. On this critical line, zero-energy modes may occur at finite momenta satisfying $k_0=\pm 2\arccos[h/(2 \mathit{\Gamma})]$.  To determine the low-energy dispersion, we set $k=k_0+q$, with $|q|\ll1$. Expanding around the gap closing momentum, we obtain
\begin{equation}
E_{+}(k_0+q)
\simeq
\sqrt{4\mathit{\Gamma}^2-h^2}|q|.
\label{eq_Ekcq}
\end{equation}
That is, the dynamical critical exponent is also $z=1$. Notably, at the boundary $|h|= 2 \mathit{\Gamma}$ [blue dot in Fig. \ref{fig_gap} (b)] of this critical line, the low-energy dispersion becomes
\begin{equation}
E_{+}(k)
\simeq
\frac{\mathit{\Gamma}}{2}k^2.
\label{eq_EkG}
\end{equation}
Here the boundary is Lifshitz-type \cite{Lifshitz1960, Wang2023Lifshitz}, because a pair of finite-momentum zero-energy nodes $k_0=\pm 2\arccos[h/(2 \mathit{\Gamma})]$ merges at $k=0$ and subsequently disappears, resulting in a quadratic critical dispersion with $z=2$.
At a $z=2$ Lifshitz critical point, the ground-state entanglement is model dependent. It may vanish for a free-fermion ground state, remain finite, or acquire anomalous zero-mode contributions, despite the presence of gapless quadratic excitations \cite{Rodney2013scaling, Hartmann2021entanglement, Wang2022universal}. We will investigate this in the next section.

\begin{figure*}[hbt]
\centering
  \includegraphics[width=1\textwidth]{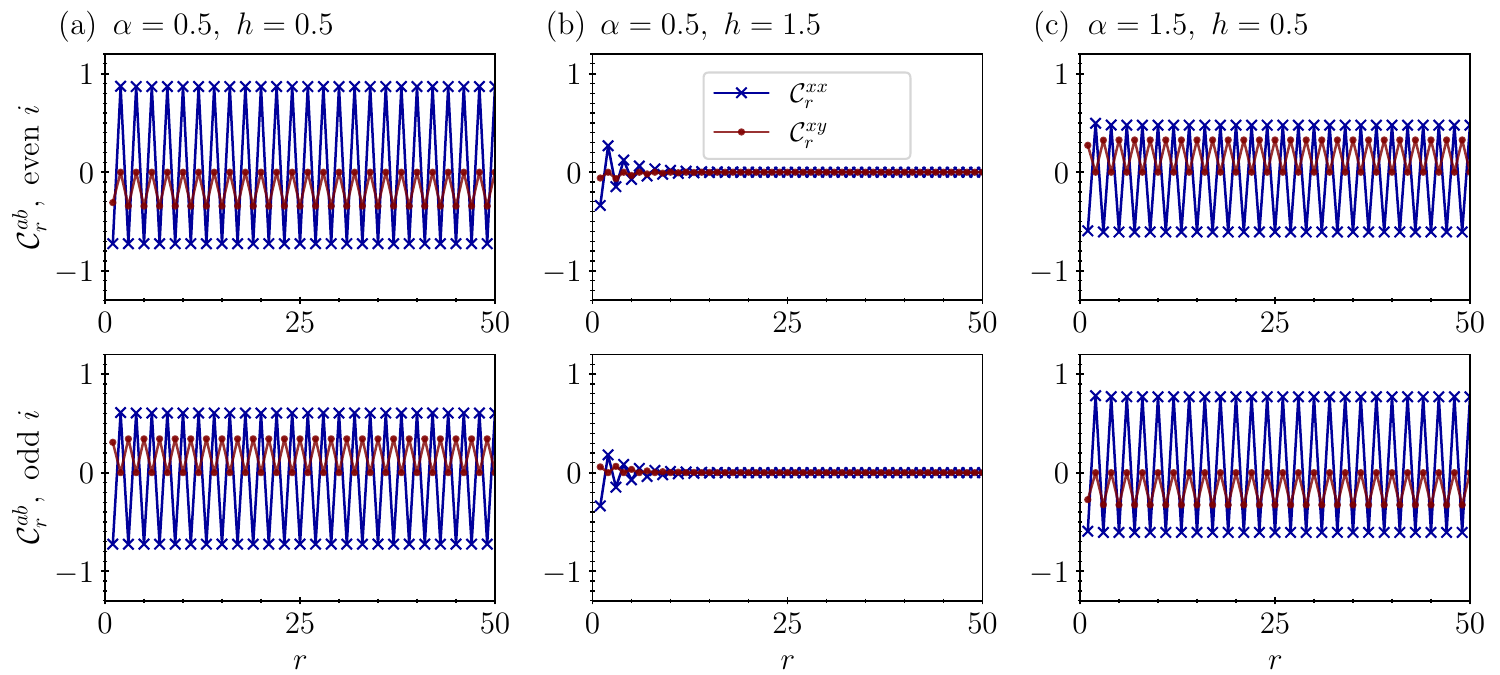}
  \caption{The correlation functions $\mathcal{C}_{r}^{xx}$ and $\mathcal{C}_{r}^{xy}$ with respect to distance $r$ for the case of $J=1$ and for different $\alpha$ and $h$. The top and bottom panels show the correlation functions for even and odd $i$ sites, respectively, with $r=j-i$. The values of $\alpha$ and $h$ are marked in Fig. \ref{fig_gap}(a) for each panels: left panel (a) $\alpha=0.5,\ h=0.5$; middle panel (b) $\alpha=0.5,\ h=1.5$ and right panel  (c) $\alpha=1.5,\ h=0.5$. Other parameters are taken as $\mathit{\Gamma}=0.8$ and $N=1000$.}
  \label{fig_correlation_J1}
\end{figure*}

\begin{figure*}[hbt]
\centering
  \includegraphics[width=1\textwidth]{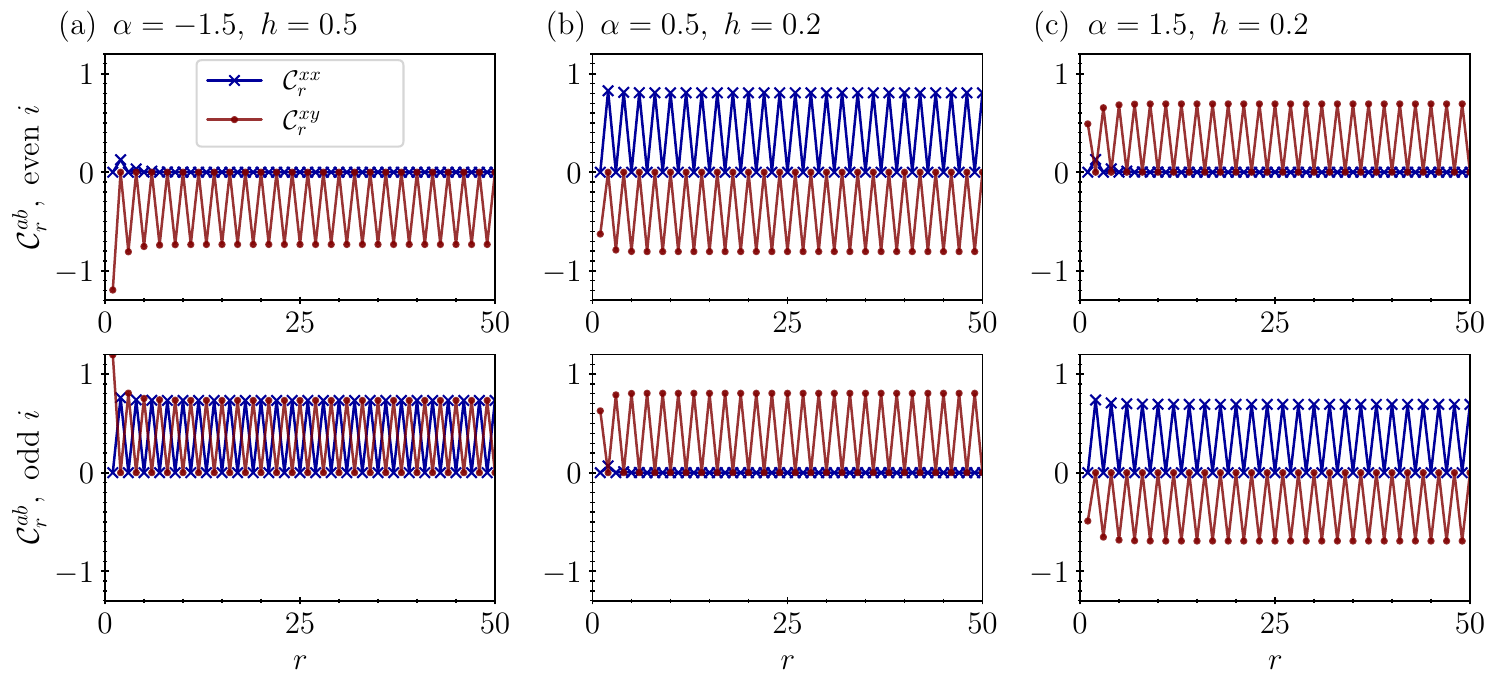}
  \caption{The correlation functions $\mathcal{C}_{r}^{xx}$ and $\mathcal{C}_{r}^{xy}$ with respect to distance $r$ for the case of $J=0$ and for different $\alpha$ and $h$. The top and bottom panels show the correlation functions for even and odd $i$ sites, respectively, with $r=j-i$. The values of $\alpha$ and $h$ are marked in Fig. \ref{fig_gap}(b) for each panels: left panel (a) $\alpha=-1.5,\ h=0.5$; middle panel (b) $\alpha=0.5,\ h=0.2$ and right panel  (c) $\alpha=1.5,\ h=0.2$. Other parameters are taken as $\mathit{\Gamma}=0.8$ and $N=1000$.}
  \label{fig_correlation_J0}
\end{figure*}

\section{Spin Correlations and entanglement entropy}
\label{sec_correlations}

In this section, we investigate the behavior of the spin-spin correlation functions and subsystem entanglement entropy in different regions of the phase diagram, including at the critical lines. The exact ground state of the quadratic fermionic Hamiltonian Eq. (\ref{eq_Hf}) is a Gaussian state, which allows us to evaluate the spin-spin correlation functions from the Majorana two-point correlation functions by means of Wick's theorem \cite{wick1950evaluation, Surace2022Fermionic}. 

\subsection{Spin-spin correlation functions}

Insights into the nature of the ground state $ |\mathrm{GS}\rangle$ can be obtained from the connected correlation functions. Here we consider the spin-$xx$ correlation function
\begin{equation}
	\mathcal{C}_{r}^{xx} = \langle \hat{\sigma}_{i}^{x} \hat{\sigma}_{j}^{x} \rangle - \langle \hat{\sigma}_{i}^{x} \rangle \langle \hat{\sigma}_{j}^{x} \rangle,
\end{equation}
and the antisymmetric connected spin correlation function in the $xy$ components
\begin{equation}
	\mathcal{C}_{r}^{xy} = \left(\langle \hat{\sigma}_{i}^{x} \hat{\sigma}_{j}^{y} \rangle-\langle \hat{\sigma}_{i}^{x} \rangle \langle \hat{\sigma}_{j}^{y} \rangle\right) -\left( \langle \hat{\sigma}_{i}^{y} \hat{\sigma}_{j}^{x} \rangle - \langle \hat{\sigma}_{i}^{y} \rangle \langle \hat{\sigma}_{j}^{x} \rangle\right),
\end{equation}
which can also be interpreted as the connected part of the $z$ component of the vector spin chirality between sites $i$ and $j$ \cite{Hikihara2001}. Here we assume $i<j$ and introduce $r=j-i$. Because the Hamiltonian has a two-site unit cell, the correlation functions generally depend not only on the distance $r$, but also on whether the
initial site $i$ belongs to the odd or even sublattice. 

The numerical results in Fig. \ref{fig_correlation_J1} compares the spatial dependence of the connected correlation functions $\mathcal{C}_{r}^{xx}$ and $\mathcal{C}_{r}^{xy}$ for different values of $\alpha$ and $h$, in the presence of Ising interaction $J=1$. The locations of $\alpha$ and $h$ in the phase diagram for the cases of Figs. \ref{fig_correlation_J1} (a), (b) and (c) are marked as red points  a, b and c, respectively, in Fig. \ref{fig_gap}(a).  
At $h=0.5$, both $\alpha=0.5$ and $\alpha=1.5$ exhibit long-range, period-two oscillations. In particular, $\mathcal{C}_r^{xx}$ alternates between positive values at even separations and negative values at odd separations, with an essentially distance-independent amplitude, consistent with long-range staggered transverse spin correlations. However, the antisymmetric connected correlation $\mathcal{C}_r^{xy}$ displays a pronounced parity dependence: it is zero for even $r$, whereas it remains finite for odd $r$, with opposite signs for even and odd reference sites $i$. Increasing $\alpha$ from $0.5$ to $1.5$ at fixed $h=0.5$ reverses this sublattice-dependent sign pattern and redistributes the amplitudes of $\mathcal{C}_r^{xx}$ between the two reference-site sublattices, while preserving the nondecaying character of the correlations. By contrast, at $\alpha=0.5$ and $h=1.5$, both correlation functions rapidly approach zero with increasing separation for either parity of $i$, leaving only short-range correlations. These results demonstrate a clear field-induced suppression of the persistent transverse and antisymmetric spin correlations observed at the lower field.

Figure \ref{fig_correlation_J0} shows the spatial dependence of $\mathcal{C}_r^{xx}$ and $\mathcal{C}_r^{xy}$ for $J=0$. The locations of $\alpha$ and $h$ in the phase diagram for the cases of Figs. \ref{fig_correlation_J0} (a), (b) and (c) are marked as red points  a, b and c, respectively, in Fig. \ref{fig_gap}(b).  For all three parameter sets, $\mathcal{C}_r^{xy}$ is zero at even separations but approaches a finite, distance-independent value at odd separations after an initial short-distance variation. Its sign is opposite for even and odd reference sites: it is negative for even $i$ and positive for odd $i$ in panels (a) and (b), whereas this pattern is reversed in panel (c). In contrast, the persistent spin-$xx$ correlations are confined to one sublattice. In panels (a) and (c), $\mathcal{C}_r^{xx}$ remains positive at even separations for odd reference sites, while the corresponding correlations for even reference sites rapidly decay toward zero. Panel (b) exhibits the opposite sublattice selection, with persistent spin-$xx$ correlations occurring for even reference sites. 

\begin{figure}[tb]
\centering
  \includegraphics[width=0.5\textwidth]{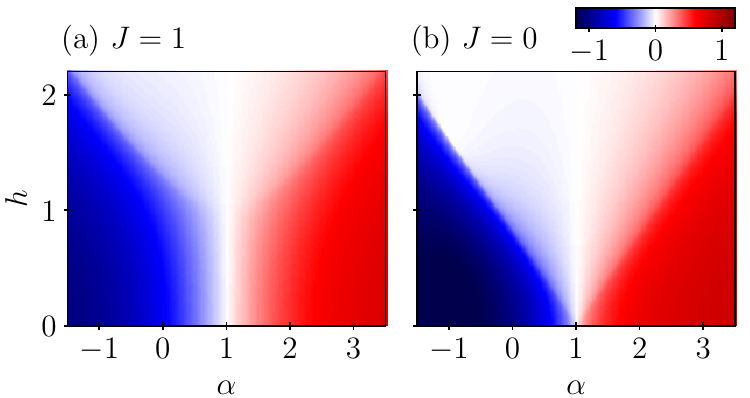}
  \caption{The contour map of the staggered chiral order parameter $\mathcal{O}_{\mathrm{stag}}^{xy}$ as a function of $\alpha$ and $h$. (a) $J=1$. (b) $J=0$. In both cases we take $\mathit{\Gamma}=0.8$, $i=100$ and $N=200$.}
  \label{fig_Qxy}
\end{figure}

Since $\mathcal{C}_r^{xy}$ at odd separations changes sign between even and odd reference sites $i$, we introduce the staggered chiral order parameter \cite{Soerensen2023, Abbasi2025}    
\begin{equation}
	\mathcal{O}_{\mathrm{stag}}^{xy}=(-1)^i \mathcal{C}_{r=1}^{xy}
\end{equation}
to characterize the quantum phase in the $\alpha$-$h$ plane. The numerical results are presented in Figs. \ref{fig_Qxy} (a) and (b) for the cases of $J=1$ and $J=0$, respectively.
In both cases, $\mathcal{O}_{\mathrm{stag}}^{xy}$ changes from negative to positive across $\alpha=1$, where it nearly vanishes, indicating a reversal of the staggered chirality. The pronounced variations in the chiral response broadly follow the gap-closing boundaries in Fig.~\ref{fig_gap}, although the maps alone cannot establish their singular nature. The sign-reversal line $\alpha=1$, however, does not generally coincide with a gap-closing boundary, and the additional critical line for $J=0$ has no clearly resolved signature in the chiral map. Thus, $\mathcal{O}_{\mathrm{stag}}^{xy}$ captures the orientation and strength of staggered chirality but cannot distinguish all regions separated by gap-closing lines. To characterize the quantum phases more fully, we next examine correlation functions and entanglement entropy at the critical points, together with the DQPTs in the next section.

\begin{figure}[tb]
\centering
  \includegraphics[width=0.5\textwidth]{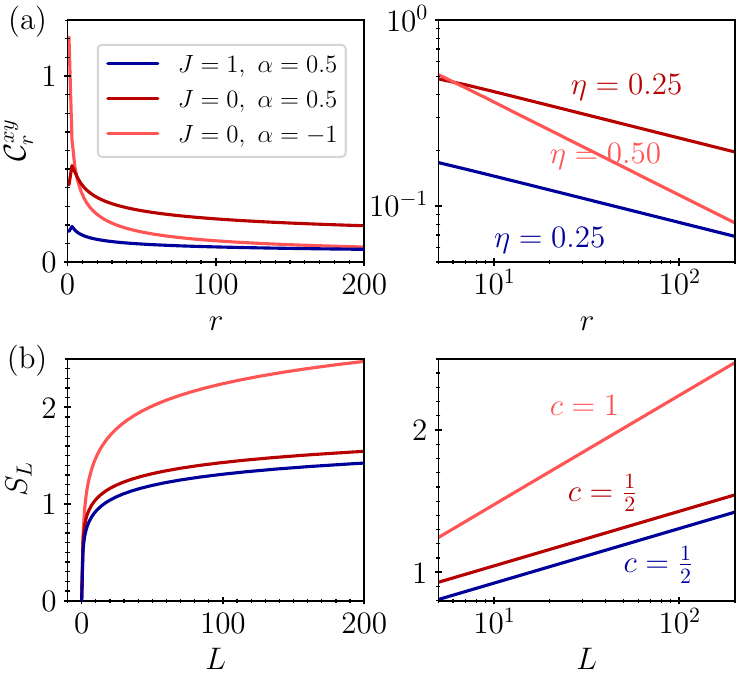}
  \caption{(a) Correlation function $\mathcal{C}_{r}^{xy}$ with respect to distance $r$; (b) Subsystem entanglement entropy $S_{L}$ with respect to  subsystem size $L$ for three representative critical points: (i) $J=1$, $\alpha=0.5$, $h_c= \sqrt{J^2+(1-\alpha)^2\mathit{\Gamma}^2}$; (ii) $J=0$, $\alpha=0.5$, $h_c= \sqrt{J^2+(1-\alpha)^2\mathit{\Gamma}^2}$; and (iii) $J=0$, $\alpha=-1$, $h=0.5$. The right panel shows $\mathcal{C}_{r}^{xy}$ and $S_{L}$ in double-logarithmic scale and semi-logarithmic scale, respectively. Other parameters are taken as $\mathit{\Gamma}=0.8$ and $N=5000$.}
  \label{fig_Cxy_SL}
\end{figure}

\subsection{Quantum criticality}

In Fig. \ref{fig_Cxy_SL}, we present the numerical results of correlation function $\mathcal{C}_{r}^{xy}$ and subsystem entanglement entropy $S_{L}$ at three representative critical points. Here the subsystem entanglement entropy is defined as 
\begin{equation}
	S_{L} = -\mathrm{Tr}\left(\hat{\rho}_{L}\ln \hat{\rho}_{L}\right) 
\end{equation}
with
\begin{equation}
	\hat{\rho}_{L}=\mathrm{Tr}_\mathcal{B} \left( |\mathrm{GS}\rangle \langle \mathrm{GS}| \right)
\end{equation}
being the reduced density matrix of the subsystem $\mathcal{A}$ of size $L$, and $\mathcal{B}$ being the complement.  As shown in Fig.~\ref{fig_Cxy_SL}(a), the vector spin chirality exhibits an algebraic decay at all three critical points,
\begin{equation}
    \mathcal{C}_{r}^{xy} \sim A r^{-\eta},
\end{equation}
where $A$ is a nonuniversal amplitude. For $\alpha=0.5$, both $J=1$ and $J=0$ yield $\eta \simeq 1/4$, although the correlation amplitude is larger for $J=0$. In contrast, the critical point at $J=0$, $\alpha=-1$ and $h=0.5$ exhibits a faster decay with $\eta \simeq 1/2$. The entanglement entropy provides a complementary distinction between these critical behaviors. As shown in Fig.~\ref{fig_Cxy_SL}(b), $S_L$ grows logarithmically with the subsystem size. For a bulk interval in the regime $1\ll L\ll N$, the conformal scaling takes the form
\begin{equation}
    S_L = \frac{c}{3}\ln L+s_0,
\end{equation}
where $c$ is the central charge and $s_0$ is a nonuniversal constant \cite{Calabrese2004}. The extracted central charges are $c\simeq 1/2$ for both critical points at $\alpha=0.5$, but $c\simeq 1$ for $J=0$, $\alpha=-1$ and $h=0.5$. The central charge $c\simeq 1/2$ at the two critical points with $\alpha=0.5$ is consistent with Ising criticality. However, at the critical line $\alpha=-1$ in the absence of Ising interaction, the excitation spectrum exhibits two gapless nodes at $k=\pm k_0$, with linear dispersion in their vicinity [see Eq. (\ref{eq_Ekcq})]. Together with the central charge $c\simeq 1$, this is consistent with a massless Dirac low-energy description.

The behavior of the ground state and its subsystem entanglement entropy at the intersection of the above two critical lines for the system with $J=0$ is of particular interest.  Consistent with the zero-energy mode at $k=0$ discussed in Eq. (\ref{eq_EkG}), the ground-state manifold at $J=0$ and  $h=2\mathit{\Gamma}$ is twofold degenerate. It can be checked that the two ground states have the following simple forms 
\begin{equation}
\begin{aligned}
|\Psi_0\rangle
&=\prod_{i=1}^{N} \left\vert \downarrow \right>_{i},\\
|\Psi_1\rangle
&=\frac{1}{\sqrt{N}}
\sum_{j=1}^{N/2}
\left(\hat{\sigma}_{2j-1}^{+}+\ii \hat{\sigma}_{2j}^{+}\right)
|\Psi_0\rangle,
\end{aligned}
\end{equation}
where $\hat{\sigma}_j^{+}=(\hat{\sigma}_j^x+\ii\hat{\sigma}_j^y)/2$ and
$\hat{\sigma}_j^z\left\vert \downarrow \right>_{j}=-\left\vert \downarrow \right>_{j}$.
The second state represents a delocalized single-spin-flip excitation in the two-site unit cell. Both states have the energy
\begin{equation}
E_0=E_1=-Nh=-2N\mathit{\Gamma}.
\end{equation}
These results show that the closing of the excitation gap at this $z=2$ quantum critical point does not necessarily imply a divergent or even nonzero ground-state entanglement entropy. The zero-energy mode allows a fully polarized product state and an entangled single-spin-flip state to coexist within the ground-state manifold, with half-chain entanglement entropies of $0$ and $\ln 2$, respectively. Thus, the quadratic gapless spectrum characterizes the critical low-energy excitations, whereas the ground-state entanglement depends on the occupation of the zero mode and the choice of state within the degenerate manifold. 

\section{Dynamical quantum phase transitions}
\label{sec_DQPTs}

The phase diagram obtained from the staggered chiral order parameter in Fig. \ref{fig_Qxy} does not show a one-to-one correspondence with that inferred from the energy gap in Fig. \ref{fig_gap}. This motivates us to investigate DQPTs as a complementary probe of the model. In particular, the four-band structure of the $k$-space BdG Hamiltonian raises the question of how different bands contribute to the quench dynamics. 

\begin{figure*}[hbt]
\centering
  \includegraphics[width=1\textwidth]{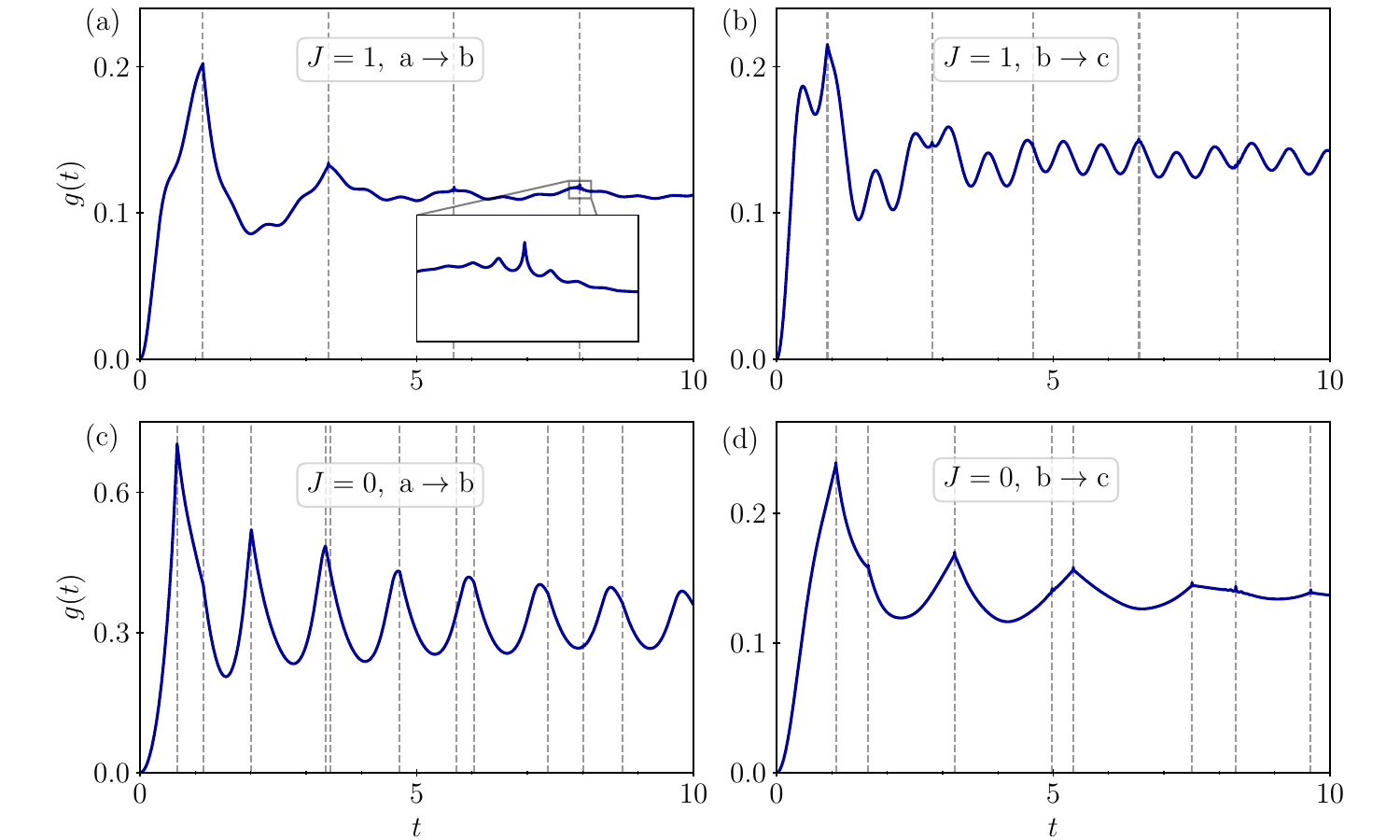}
  \caption{Rate functions for different pre-quench and post-quench Hamiltonians following a sudden quench. The corresponding quench paths for top and bottom panels  are marked in Figs. \ref{fig_gap} (a) and (b), with $J=1$ and $J=0$, respectively. Specifically, the parameters $(\alpha,\ h)$ for pre-quench and post-quench Hamiltonians are taken as (a) $(0.5,\ 0.5)\rightarrow (0.5,\ 1.5)$; (b) $(0.5,\ 1.5)\rightarrow (1.5,\ 0.5)$; (c) $(-1.5,\ 0.5)\rightarrow (0.5,\ 0.2)$; (c) $(0.5,\ 0.2)\rightarrow (1.5,\ 0.2)$. The gray dashed lines indicate the critical time $t_{\mathrm{c}}$ of DQPTs solved from Eq. (\ref{eq_critical_eq}), and the inset provides a magnified view of a cusp. Other parameters are taken as $\mathit{\Gamma}=0.8$ and $N=2000$.}
  \label{fig_ratefun}
\end{figure*}

\subsection{Critical condition for DQPTs}
Assuming that the system is initially in its ground state $ \lvert\mathrm{GS}\rangle$, it evolves under the post-quench Hamiltonian $\hat{H}^{f}=\sum_{k}\hat{\mathcal{H}}^{f}_{\mathrm{BdG}}(k)$. 
Then the Loschmidt amplitude of the system can be factorized over independent momentum sectors
\begin{equation}
\mathcal{G}(t)=\prod_k \mathcal{G}_k(t), 
\end{equation}
since the lattice momentum is conserved.
Introducing the notation of the ground state $ \lvert\mathrm{GS}\rangle = \prod_{k>0} \prod_{\lambda=\pm}| \psi_{k,\lambda}^{-} \rangle$ with $| \psi_{k,\lambda}^{-} \rangle=\gamma_{k,\lambda}\gamma_{-k,\lambda} | 0 \rangle$ and $| \psi_{k,\lambda}^{+} \rangle=\gamma_{k,\lambda}^{\dagger}\gamma_{-k,\lambda}^{\dagger} | 0 \rangle$, the momentum-resolved Loschmidt factor can be expressed as \cite{Mendl2019}
\begin{equation}
\mathcal{G}_k(t)=\det\!\left[\left< \psi_{k,\lambda_{1}}^{-,i} \right\vert e^{-\mathrm{i} \hat{\mathcal{H}}^{f}_{\mathrm{BdG}} \left( k \right) t} \left\vert  \psi_{k,\lambda_{2}}^{-,i} \right>
\right]_{\lambda_1, \lambda_2 =\pm},
\end{equation}
The superscripts $i$ and $f$ denote the corresponding physical quantities before and after the quench, respectively. Note that $\mathcal{G}_k(t)$ can be written ‌explicit‌ly by inserting the completeness relation of the eigenstates of the post-quench Hamiltonian. By doing this, we obtain
\begin{equation}
\mathcal{G}_k(t)=
\det\!\left[
(Q^{-,i}_{k})^{\dagger}\mathcal{Q}^{f}_{k}
e^{-\mathrm{i}\mathcal{D}^{f}_{k}t}
(\mathcal{Q}^{f}_{k})^{\dagger}Q^{-,i}_{k}
\right].
\end{equation}
Here, $(\mathcal{Q}^{f}_{k})^{\dagger}Q^{-,i}_{k}$ is a $4\times2$ matrix that incorporates both the change in the Bogoliubov coefficients and the mixing between the $\lambda=\pm$ branches, and the columns of $\mathcal{Q}_k^{f}=(Q_k^{-,f},Q_k^{+,f})$ are the eigenvectors of the post-quanch BdG Hamiltonian in each $k>0$ subspace under the basis of  $\Psi_k$, with the corresponding diagonal eigenenergy matrix 
\begin{equation}
\mathcal{D}_k^{f}
=
\operatorname{diag}
\left(
-E_+^{f},-E_-^{f},E_+^{f},E_-^{f}
\right).
\end{equation}

In fact, the eigenvectors $Q_k^{\pm}$ can be expressed as $Q_k^{\pm}= T_k \widetilde{Q}_k^{\pm}$, where $\widetilde{Q}_k^{\pm}$ are the eigenvectors under the basis of $\widetilde{\Psi}_k=(c_{k,+},\ c_{k,-},\ c^\dagger_{-k,+},\ c^\dagger_{-k,-})^{T}$. The negative and positive energy branches of  eigenvectors are
\begin{equation}
\widetilde{Q}_k^-=
\begin{pmatrix}
i v_{k,+} & 0\\
0 & -i v_{k,-}\\
u_{k,+} & 0\\
0 & u_{k,-}
\end{pmatrix},
\end{equation}
and
\begin{equation}
\widetilde{Q}_k^+=
\begin{pmatrix}
u_{k,+} & 0\\
0 & u_{k,-}\\
i v_{k,+} & 0\\
0 & -i v_{k,-}
\end{pmatrix}, 
\end{equation}
respectively, and the transformation matrix is
\begin{equation}
	T_k=
\begin{pmatrix}
U_k & 0\\
0 & U_{-k}^{*}
\end{pmatrix},\quad U_k=
\frac{1}{\sqrt{2}}
\begin{pmatrix}
e^{i\chi} & e^{i\chi}\\
i e^{i\beta_k} & -i e^{i\beta_k}
\end{pmatrix}.
\label{eq_T_U}
\end{equation}
Here, we suppress the superscripts $i$ and $f$ for notational simplicity.

Analogous to a free-energy density in equilibrium, the rate function 
\begin{equation}
	g\left( t \right) =-\frac{1}{N} \sum_{k} \ln \left| \mathcal{G}_{k} \left( t \right) \right|
\end{equation}
exhibits nonanalyticities at critical times associated with the DQPTs. In Fig. \ref{fig_ratefun}, we show the numerical results of the rate functions for different pre-quench and post-quench Hamiltonians following a sudden quench. The corresponding quench paths for top and bottom panels are marked by the arrows in Figs. \ref{fig_gap} (a) and (b), with $J=1$ and $J=0$, respectively.
The critical momentum $k_{\mathrm{c}}$ and critical time $t_{\mathrm{c}}$ of the DQPTs are determined by the condition of Loschmidt zeros 
\begin{equation}
	\begin{cases}
	\operatorname{Re}\mathcal{G}_{k_{\mathrm{c}}}(t_{\mathrm{c}})=0,\\ \operatorname{Im}\mathcal{G}_{k_{\mathrm{c}}}(t_{\mathrm{c}})=0.
	\end{cases}
	\label{eq_critical_eq}
\end{equation}
We numerically solve this equation for the cases in Fig. \ref{fig_ratefun} and plot the the critical time $t_{\mathrm{c}}$ of the DQPTs with gray dashed lines, which is consistent with the cusps of their corresponding rate functions. 

\subsection{Criterion for branch mixing}

The numerical results in Fig. \ref{fig_ratefun} indicate that the critical times of DQPTs form periodic sequence in Fig. \ref{fig_ratefun} (a) and aperiodic sequences in Figs. \ref{fig_ratefun} (b)-(d). Equally spaced critical times typically occur in two-band quenches with a time-independent post-quench Hamiltonian, when a single critical momentum has equal populations in the two final bands. In four-band models, critical times are generally not equally spaced due to multiband interference \cite{cao2024aperiodic}. For the model considered in this paper, we show that the  equal spacing of critical time can nevertheless arise when the Loschmidt zeros are governed by a single effective two-level sector at a fixed critical momentum in the following.

The overlap matrix between the pre-quench and post-quench branch bases
is
\begin{equation}
S_k=(U_{k}^{f})^{\dagger}U_{k}^{i},
\end{equation}
where $U_{k}^{i/f}$ is defined in Eq. (\ref{eq_T_U}). Then the interbranch mixing probability is
\begin{equation}
\left|(S_k)_{+,-}\right|^2
=
\sin^2\!\left(
\frac{\phi_i-\phi_f}{2}
\right).
\end{equation}
Therefore, if $\phi_i\neq\phi_f$, the two branches are mixed. 
    While for $\phi_i=\phi_f$, we have $\left|(S_k)_{+,-}\right|^2=0$ and there is no branch mixing, and the
    problem reduces to two independent $2\times2$ sectors. This is satisfied when the quench involves only a change in the field from $h_i$ to $h_f$, while $J$, $\alpha$, and $\mathit{\Gamma}$ remain fixed [see the definition of $\phi_{i/f}$ in Eq. (\ref{eq_phi_theta})]. In this case, the Loschmidt amplitude can be factorized as 
    \begin{equation}
    	\mathcal{G}(t) = \prod_{k} \prod_{\lambda=\pm 1} \mathcal{G}_{k,\lambda}(t),
    \end{equation}
    with
    \begin{equation}
    	\mathcal{G}_{k,\lambda}(t) = \cos \left[ E_{\lambda}^{f}(k) t \right] + i \left(1 - 2 p_{k,\lambda}\right) \sin \left[ E_{\lambda}^{f}(k) t \right],
    \end{equation}
    and $p_{k,\lambda} = (u_{k,\lambda}^{f} v_{k,\lambda}^{i} - v_{k,\lambda}^{f} u_{k,\lambda}^{i})^{2}$. The critical momentum is determined by 
    \begin{equation}
    	1 - 2p_{k_{c},\lambda} = 0,
    \end{equation}
    and the equally spaced critical time is
    \begin{equation}
    	t_{c,n}^{(\lambda)} = \frac{(2n+1)\pi}{2E_{\lambda}^{f}(k_c)}, \quad n = 0, 1, 2, \ldots .
    \end{equation}
This is in accord with the numerical results  in Fig. \ref{fig_ratefun} (a). 
 
 Notably, the cusps of rate function in Fig. \ref{fig_ratefun} (c) corroborate that the parameters at points a and b in Fig. \ref{fig_gap} (b) correspond to different phases, which is not reflected by the staggered chiral order parameter in Fig. \ref{fig_Qxy} (b) while is captured by the spin-$xx$ correlation in Figs. \ref{fig_correlation_J0} (a) and (b). Thus, the occurrence of DPQTs corresponds to transitions in the spin-$xx$ correlations, but does not exhibit a one-to-one correspondence with transitions in vector spin chirality.

\section{Conclusion}

\label{sec_conclusion} 

In summary, our study of the transverse-field Ising chain with dimerized anisotropic Gamma interactions reveals how the competition between Gamma anisotropy and the transverse field shapes equilibrium order and nonequilibrium criticality, both with and without Ising interactions. Spin-spin correlations and the vector chiral order parameter distinguish positive and negative vector chiral phases from a paramagnetic phase. In the absence of Ising interactions, we identify a gapless critical line with no observed finite-size drift, where spin correlations and subsystem entanglement entropy exhibit behavior distinct from Ising criticality. We further show that the quench path determines whether DQPT critical times form periodic or aperiodic sequences. Together, these results establish dimerized off-diagonal exchange as a mechanism for enriching both the equilibrium phase structure and the dynamical critical behavior of quantum Ising chains.

\acknowledgments This work was supported by the National Natural Science
Foundation of China (Grants No. 12505015, No. 12404493), the Guangdong Basic and Applied Basic Research Foundation (Grants No. 2024A1515110222).

%


\end{document}